\documentclass{blois}

\def\Journal#1#2#3#4{{#1} {\bf #2}, #3 (#4)}

\def\PRD{{\em Phys. Rev.} D}

\def\JHEP{\em JHEP}
\def\JINST{\em JINST}
\def\APPB{{\em Acta Phys. Polon.} B}

\def\be{\begin{equation}}
\def\ee{\end{equation}}
\def\bea{\begin{eqnarray}}
\def\eea{\end{eqnarray}}

\newcommand{\Photo}{}

\begin{document}
\vspace*{4cm}

\title{Search for the standard model deviations in top quark precision studies at CMS}
\author{K. Skovpen on behalf of the CMS collaboration}

\address{Vrije Universiteit Brussel, IIHE, Pleinlaan 2, 1050 Brussels, Belgium}

\maketitle\abstracts{
Precision studies of top quark properties provide a unique playground
to test the predictions of the standard model and to search for new
physics. Reviewed results from the CMS experiment done with the data collected at 8 TeV 
include studies of top quark Wtb anomalous and FCNC couplings,
polarization, CP-violation, and spin correlation effects. No significant deviations from the SM predictions are observed.
}

The Large Hadron Collider (LHC) is a top quark
factory that produces a large number of events containing
top quarks. The top quark takes a special place in the standard
model (SM) for numerous reasons. Top quarks can manifest themselves in
distinctive experimental signatures as top quark almost
exclusively decays to a W boson and a b quark with a very short lifetime
escaping forming any bound states. This allows experimentalists to 
directly study the top quark properties by analyzing its decay products. The fact that
this particle is the heaviest elementary particle ever discovered
additionally suggests an enhanced sensitivity to various new particles
and interactions proposed in beyond the SM (BSM) theories. New physics
phenomena can be probed in the measurement of production rates of
top quarks, in the study of the Wtb vertex structure and in the search for
interactions that are heavily suppressed in SM.

A large number of top quark physics analyses have been done
at CMS~\cite{CMS} including the measurement of the production rates of single top quark, as well as top quark
pair and top quark associative production with W, Z, Higgs bosons
and photon. The Cabibbo-Kobayashi-Maskawa (CKM) matrix element $|V_{tb}|$ is
directly extracted from the measured single top quark cross
section. All these results are in remarkable
agreement with theory predictions.

Events with single top quarks provide a unique playground to search
for deviations from the SM by studying the Wtb vertex structure, as
the single top production cross section is proportional to the strength of
Wtb interaction. The general lagrangian that defines Wtb interaction
includes several couplings that can be extracted from the measured
kinematic distributions of top quarks and its
decay products. The lagrangian includes two vector and two tensor
couplings:

\begin{equation}
\mathcal{L} =
\frac{g}{\sqrt{2}}\bar{b}\gamma^{\mu}(f_{V}^{L}P_{L}+f_{V}^{R}P_{R})tW_{\mu}^{-}-
\frac{g}{\sqrt{2}}\bar{b}\frac{\sigma^{\mu\nu}\partial_{\nu}W_{\mu}^{-}}{M_{W}}(f_{T}^{L}P_{L}+f_{T}^{R}P_{R})t,
\end{equation}

\noindent where $P_{L,R} = (1\mp\gamma_{5})/2$, $\sigma_{\mu\nu} =
i(\gamma_{\mu}\gamma_{\nu}-\gamma_{\nu}\gamma_{\mu})$, g is the
coupling constant of the weak interaction, $f_{V}^{L} (f_{V}^{R})$
represents the left-handed (right-handed) vector coupling and
$f_{T}^{L} (f_{T}^{R})$ is the left-handed (right-handed) tensor
coupling. All couplings vanish at leading-order (LO) in SM but 
$f_{V}^{L} = V_{tb}$.

The analysis that probes anomalous Wtb couplings in t-channel
production mode exploits Bayesian neural network (BNN) technique
to define two neural networks: to suppress QCD background (Multijet BNN) and
to extract t channel events from data (SM BNN)~\cite{BNN}. Additional anomalous Wtb BNNs are used
to separate individual anomalous couplings contributions from the
left-handed vector coupling component. In the limit setting procedure several
strategies are considered depending on the number of probed anomalous
couplings. A multi-dimensional extraction of limits for anomalous
couplings is possible because of negligible interference terms present 
for different types of couplings. One-dimensional limits at 95\% CL
on each of the couplings from the multi-dimensional fit after the integration over
the rest of anomalous contributions yield $f_{V}^{L} > 0.98$,
$f_{V}^{R} < 0.16$, $f_{T}^{L} < 0.057$ and $-0.049 < f_{T}^{R} < 0.048$.
As expected in the SM, all but the left-handed vector coupling are
consistent with zero.

The measurement of W helicity fractions in events with top
quarks, where W helicity denotes the projection of W boson spin on its
momentum, has a direct link to top quark properties and is sensitive
to the real part of Wtb anomalous couplings.
Each helicity fraction is defined as the relative top quark decay
probability corresponding to a specific helicity of a W boson. 
Helicity fractions are extracted from the measurement of the $\cos(\theta^{*})$
distribution. The definition of $\cos(\theta^{*})$ corresponds to the angle between
the W boson momentum in the top quark rest frame and the momentum of
the down type decay fermion in the rest frame of the W boson.

\begin{figure}[hbtp]
  \begin{center}
  \includegraphics[width=0.5\linewidth,bb=257 19 817 576]{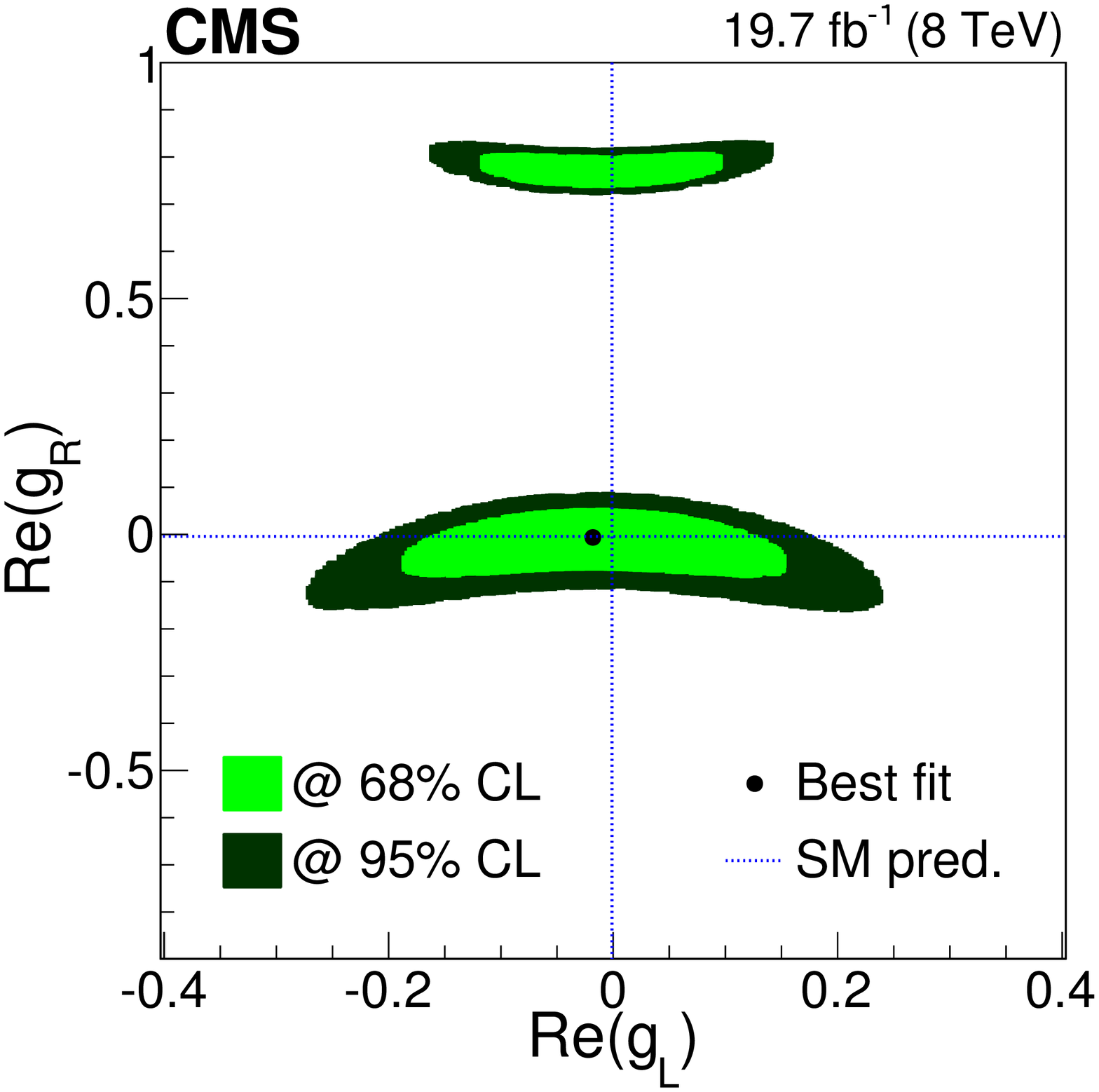}
  \caption{Exclusion limits on Wtb anomalous tensor couplings. The Wtb vector
  couplings are fixed to SM values. $f_{T}^{L} \equiv -g_{L}$,
  $f_{T}^{R} \equiv -g_{R}$.}
  \label{fig:WHEL}
  \end{center}
\end{figure}

The W boson helicity fractions can be measured in both single top and
top quark pair events. The analysis that studies
the single top production represents the first measurement of W boson helicity
done in t-channel single top events~\cite{WHEL}. Single top events are accompanied
by top quark pair events which are also considered in the analysis as
signal. The $\cos(\theta^{*})$ distribution from simulation is fitted to 
the distribution in data to extract the W boson helicity fractions. 
Left-handed and longitudinal polarizations of W boson are treated as 
free parameters in the fit, while the right-handed polarization is 
obtained from the constraint that the sum of all three polarizations must be equal 
to unity. The measured W helicities are used to set limits on the tensor terms of the
Wtb anomalous couplings by fixing the vector terms to SM predicted values.
The results are shown in Fig~\ref{fig:WHEL} and are consistent with the SM predictions.

The top quark polarization depends on the production mode and is
very small in case of pair production, while it is expected to be very
strong in case of singly produced top quarks. Deviations from SM can
be probed in the measurement of top quark polarization as various new models 
can alter the spin of the top quark. In top quark pair events one can additionally study effects arising from spin correlations of
top quarks.

The measurements of top quark polarization and spin correlations
can be re-interpreted via the presence of anomalous ttg couplings
that can significantly modify the shape of distributions of
several kinematic variables. These anomalous interactions can be probed
within the effective model of chromo-electric
($\hat{d_{t}}$) and chromo-magnetic ($\hat{\mu_{t}}$) dipole
moments~\cite{DIFF}. Spin correlations and top quark polarization
are sensitive to $Re(\hat{\mu_{t}})$ and
$Im(\hat{d_{t}})$, respectively. The value of $Re(\hat{\mu_{t}})$ is determined 
from the measurement of normalized differential cross section which can be modified in the
presence of new physics. SM and new physics contributions are
parametrized by polynomial functions which are used in template
fit to the measured normalized differential cross section.
CP-violating component of top quark polarization is sensitive to
$Im(\hat{d_{t}})$. Upper limits of $-0.053 < Re(\hat{\mu_{t}}) < 0.026$
and $-0.068 < Re(\hat{d_{t}}) < 0.067$ are set at 95\% CL.

\begin{figure}[hbtp]
  \begin{center}
  \includegraphics[width=0.5\linewidth,bb=270 171 791 566]{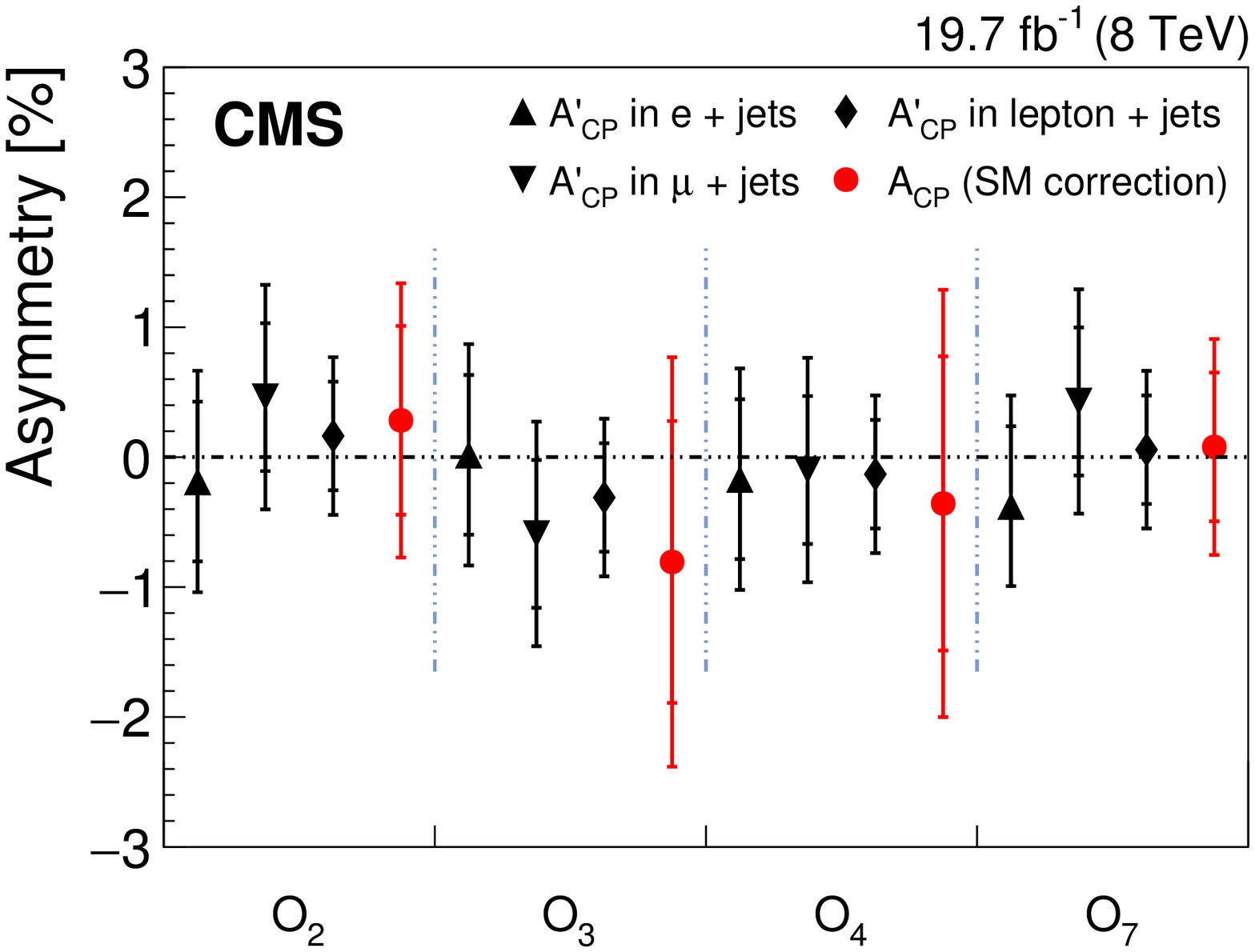}
  \caption{Summary of the uncorrected (corrected) CP asymmetries
  $A'_{CP}$ ($A_{CP}$).}
  \label{fig:CP}
  \end{center}
\end{figure}

In the SM, CP violation in production and decay of top quarks is
predicted to be very small. However, in many BSM models a sizable CP violation
effects can be observed to provide a possible explanation for the
matter-antimatter asymmetry in the Universe. A first measurement of 
CP-violating asymmetries in production and
decays of top quarks exploits T-odd, triple-product
correlations~\cite{CP}. Several
observables ($O_{i}$) are defined that represent a combination of
a reconstructed object's spin and momentum vectors. Asymmetry for each
of these observables is defined as:

\begin{equation}
A_{CP} (O_{i}) =
\frac{N_{events}(O_i>0)-N_{events}(O_i<0)}{N_{events}(O_i>0)+N_{events}(O_i<0)},
\end{equation}

\noindent and is measured with and without applying corrections for
detector effects as these additional corrections can also be affected by the BSM.
Results of the actual measurement in data are shown in Fig~\ref{fig:CP}. No
deviations from the SM predictions are observed.

Flavour-changing neutral currents (FCNC) are highly
suppressed at tree level in SM by the Glashow-Iliopoulos-Maiani (GIM) mechanism~\cite{GIM} but can be
significantly enhanced in various BSM~\cite{FCNCBSM}. FCNC with top quarks can occur
either in events with singly produced top quarks or in events with top
quark pairs where one of the top quarks decays via FCNC.

Top-gluon FCNC is probed in the single top t channel production
in association with an additional quark or gluon~\cite{TG}. This FCNC
analysis is done by
exploiting a similar BNN framework as is used in the analysis on the
Wtb anomalous couplings~\cite{BNN}. A difference with respect to the anomalous
Wtb analysis is that here additional BNNs are trained sensitive to FCNC
events. 
The observed (expected) limit at 95\% CL on the FCNC branching ratio of
top quark with up and charm-type quark is 
$B(t \rightarrow ug) < 2.0
(2.8) \cdot 10^{-5}$ and $B(t \rightarrow cg) < 4.1 (2.8) \cdot
10^{-4}$, respectively.

Top-Z FCNC is simultaneously probed in single top and top quark pair events
with the final state of three leptons~\cite{TZ}. Signal is
extracted from a simultaneous fit to the transverse mass of the
reconstructed W and to the final Boosted decision tree discriminant 
used to select signal events. Observed (expected) limit at 95\% CL on the FCNC branching
ratio of top quark with up and charm-type quark is $B(t \rightarrow uZ) < 2.2 \cdot 10^{-4} (2.7
\cdot 10^{-4})$ and $B(t \rightarrow cZ) < 4.9 \cdot 10^{-4} (1.2
\cdot 10^{-3})$, respectively.

\begin{figure}[hbtp]
  \begin{center}
  \includegraphics[width=0.7\linewidth,bb=265 70 820 571]{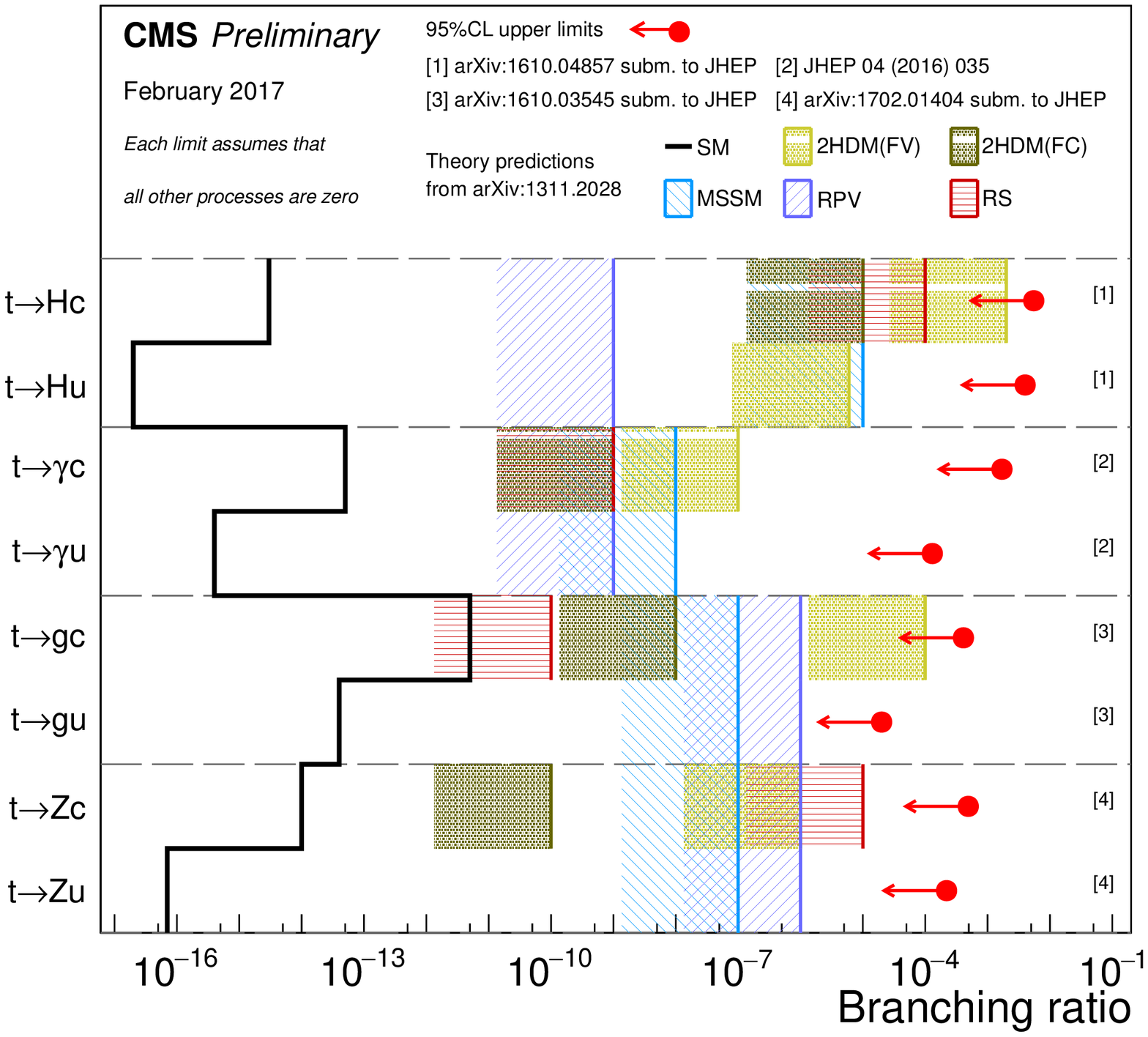}
  \caption{Summary of results on FCNC searches at CMS.}
  \label{fig:FCNC}
  \end{center}
\end{figure}

A summary of all FCNC searches at CMS is shown in
Fig~\ref{fig:FCNC}~\cite{FCNC}.

A large number of experimental results with the high precision
measurements of the top quark properties is available 
from 8 TeV analyses done with the CMS detector. 
Theoretical predictions stay in agreement
with data observations and no significant deviations from the SM
expectations are observed yet.

\end{document}